\documentclass[aps,prd,twocolumn,preprintnumbers,superscriptaddress,floatfix]{revtex4}

\usepackage{multirow, graphicx,amssymb,url,mathrsfs,amsmath}
\usepackage{eucal,wrapfig,boxedminipage,setspace,subfigure}
\usepackage{amsxtra,amstext,latexsym,dsfont}

\def\IR{{\hbox{{\rm I}\kern-.2em\hbox{\rm R}}}}
\def\IB{{\hbox{{\rm I}\kern-.2em\hbox{\rm B}}}}
\def\IN{{\hbox{{\rm I}\kern-.2em\hbox{\rm N}}}}
\def\IC{\,\,{\hbox{{\rm I}\kern-.59em\hbox{\bf C}}}}
\def\IZ{{\hbox{{\rm Z}\kern-.4em\hbox{\rm Z}}}}
\def\IP{{\hbox{{\rm I}\kern-.2em\hbox{\rm P}}}}
\def\IH{{\hbox{{\rm I}\kern-.4em\hbox{\rm H}}}}
\def\ID{{\hbox{{\rm I}\kern-.2em\hbox{\rm D}}}}

\newcommand{\beq}{\begin{equation}}
\newcommand{\eeq}{\end{equation}}
\newcommand{\bea}{\begin{eqnarray}}
\newcommand{\eea}{\end{eqnarray}}

\begin{document}

\voffset 1cm

\newcommand\sect[1]{\emph{#1}---}

\title{Scale Separation, Strong Coupling  UV Phases,  and the Identification\\ of the Edge of the Conformal Window}

\author{Anja Alfano}
\affiliation{ STAG Research Centre \&  Physics and Astronomy, University of
Southampton, Southampton, SO17 1BJ, UK}

\author{Nick Evans}
\affiliation{ STAG Research Centre \&  Physics and Astronomy, University of
Southampton, Southampton, SO17 1BJ, UK}

\begin{abstract}
We use a simple holographic model to discuss approaching the edge of the conformal window in strongly coupled gauge theories to draw lessons for lattice studies. 
Walking gauge theories have a gap between the scale where they enter the strong coupling regime and the scale of chiral symmetry breaking. We highlight that there can also be a gap between the scale where the critical value of the quark anti-quark operator's anomalous dimension is passed and the scale of the condensate. This potentially makes identifying the edge of the conformal window in a lattice simulation with UV bare coupling below the fixed point value on a finite lattice difficult. A resolution is to study the theory with a coupling above the fixed point value at the UV cut off. Here we show that an ``artefact" phase with chiral symmetry breaking triggered at the UV cut off exists and lies arbitrarily close to the fixed point at the edge of the conformal window. 
We quantify the chance of a misidentification of a chiral symmetry breaking theory as IR conformal. We also quantify where the artefact phase lies, tuned to the fixed point value. We use the latest lattice results for SU(3) gauge theory with ten quark flavours in \cite{Hasenfratz:2023wbr} as a test case; we conclude their identification that the theory is in the conformal window is reliable. 
\end{abstract}%

\maketitle

\newpage

\section{Introduction}\vspace{-0.5cm}

At large numbers of colours, $N_c$, and quark flavours, $N_f$, SU($N_c$) gauge theories are known to possess weakly coupled Banks Zak infra-red (IR) fixed points \cite{Banks:1981nn}. These fixed points lie at $N_f$ values just below where the theory gains asymptotic freedom at $N_f=11 N_c/2$ and result from a cancellation between the one- and two-loop $\beta$ function terms. A sensible story has been proposed for how these fixed points evolve into the strongly coupled regime and away from large $N_c$ \cite{Appelquist:1996dq,Dietrich:2006cm}. That account says that the IR fixed point remains as $N_f$ lowers until the fixed point is sufficiently strongly coupled that the anomalous dimension of the quark anti-quark bilinear operator has risen to one ($\gamma=1$). This criteria is natural since it is when the mass and condensate are of equal dimension and emerges in both gap equation \cite{Cohen:1988sq} and holographic analyses 
\cite{Jarvinen:2011qe,Kutasov:2011fr,Alvares:2012kr}. 

A recent lattice study of SU(3) gauge theory with $N_f=10$ \cite{Hasenfratz:2023wbr} appears to support this account, placing the theory in the conformal window. In Figure 1 we reproduce their result for the $\beta$ function of the theory as a function of $g^2$. It shows a clear fixed point around $g^2=15$. They compute that the theory has $\gamma=0.6$. It is argued that this theory lies in the strongly coupled conformal window.

Here, though, we want to play the role of devil's advocate and use that example to discuss whether such an identification could go wrong? The lattice analysis was done on a lattice that had at most $28^3\times 56$ sites and so the analysis might in principle miss large scale separations of thirty or more. We will show such a case by enacting the running that the lattice study finds in a simple holographic model of the chiral symmetry breaking. To allow ourselves a ``parameter space" we relax the condition $\gamma=1$ which is sensibly argued for but not proven. 

\begin{center}
\includegraphics[width=6.7cm,height=4.8cm]{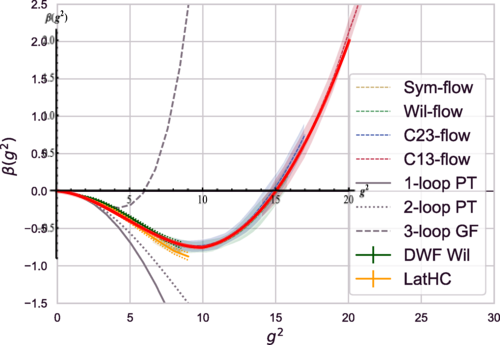}\\
\includegraphics[width=6.7cm,height=3.8cm]{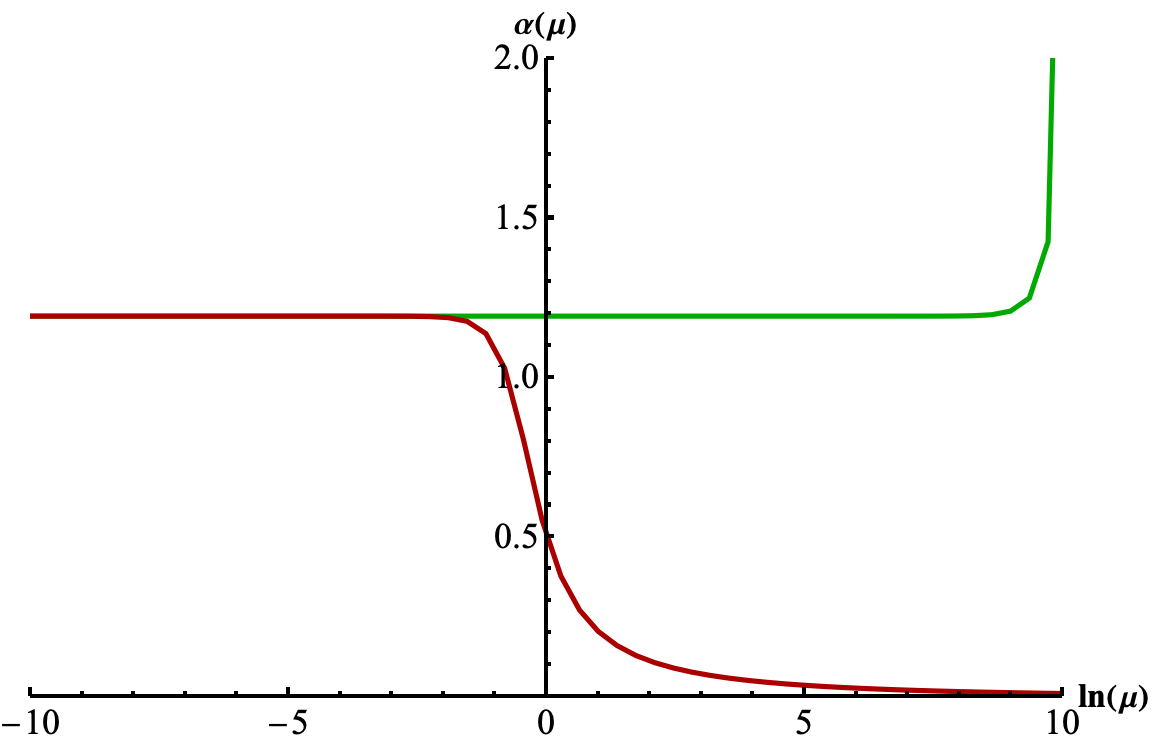}

\textit{Fig 1: The beta function plot taken from \cite{Hasenfratz:2023wbr}; our fit is overlaid in red. The resulting $\alpha(\mu)$, on integrating up the beta function, is plotted below, showing an IR fixed point at $\alpha = 15/4 \pi$. We show UV boundary conditions that approach the fixed point from above and below.}
\end{center}

In holograhy in AdS$_5$ space (with radius one) a quark condensate of dimension three, $\Delta=3$, is described by a scalar in the bulk of mass squared $-3$ (as for example, in the D3/probe D7 system \cite{Karch:2002sh,Kruczenski:2003be,Erdmenger:2007cm}). This follows from the relation $M^2=\Delta(\Delta-4)$ \cite{Witten:1998qj}. In top down chiral
symmetry breaking
 models (where supersymmetry is broken \cite{Babington:2003vm}
or with a magnetic field \cite{Albash:2007bk}) this mass squared becomes dependent on the radial (renormalization group flow) direction, decreasing into the IR \cite{Alvares:2012kr}. When it passes through $M^2=-4, \Delta=2, \gamma=1$ the Breitenlohner-Freedman (BF) bound \cite{Breitenlohner:1982jf} is violated and the scalar becomes unstable to acquiring a vacuum expectation value (vev), and chiral symmetry breaking is triggered.

This discussion though is only strictly valid in AdS$_5$, yet, generically, the space will be deformed by the relevant perturbation causing the chiral symmetry breaking. It is therefore not proven that when the BF bound is violated in a deformed space, that $\gamma=1$. Thus below, although we will use a fixed AdS$_5$ geometry, we will only assume that the deviation of the scalar mass from $-3$ ($\Delta m^2$) is proportional to the running $\alpha$. We will study the behaviour of the theory as a function of the proportionality constant $k$. We will study the separation in scale between where the BF bound is violated and the chiral condensate forms. If a theory is sufficiently walking and just barely above the BF bound violation point in the IR then a big gap between these scales is possible. This is simply an energetic argument - if the energy cost of ``living with" the instability is small it can take a spread of scales for the benefit of allowing condensation to be felt.  We find a separation of scales greater than 35 if $\Delta m^2$ lies between 1-1.3 of its critical value.

So, do we trust the lattice assignment \cite{Hasenfratz:2023wbr} of the $N_f=10$ theory to the conformal window? If the reader believes a strict $\gamma=1$ criteria, then yes. If one instead believes that the critical value might lie between $\gamma=0.4$-1.4 then the measured value of 0.65 would be a concern if the true critical value were between 0.5-0.65, i.e. $15\%$ of the range. 
The lattice work \cite{Hasenfratz:2023wbr} seeks to resolve this by studying the theory with a choice of bare coupling at the UV cut off (lattice scale) that lies above the fixed point value. These theories have a UV Landau pole in the continuum, but the lattice UV cut off can protect the IR from knowledge of its existence. They observed no chiral symmetry breaking even with the UV coupling 30\% or so above the fixed point value, and this indeed suggests that the fixed point preserves chiral symmetry. This analysis is a crucial and convincing aspect of their work. 

The second part of our work is to ask: can this method of studying the theory at couplings above the critical coupling always be easily used? As the coupling is raised in the UV in the holographic model, it will at some point violate the BF bound at the UV scale. If it does, then there is no stable vacuum in the holographic model. We associate this phase to the ``lattice artefact phase" often seen in lattice simulations if the bare coupling is taken too large \cite{Hasenfratz:2021zsl}. It is associated with chiral symmetry breaking at the lattice scale (and is also not rotationally invariant). In particular we point out that as the fixed point coupling value moves towards the BF bound violating value, this artefact phase moves very close to the physical phase. A lattice simulation would need to fine tune the coupling to the fixed point value, from above, to be fully certain that the theory is in the conformal window.

Our conclusion is that, were a theory to lie within 10\% (for example in $N_f$) of the edge of the conformal window, it would be very difficult to identify the phase. On a finite lattice, chiral symmetry breaking theories may look conformal on the lattice when the UV coupling is picked below the fixed point value. On the other hand, picking UV couplings above the fixed point will likely leave a theory in the artefact phase. The resolution of the behaviour becomes arbitrarily harder the closer one approaches the edge of the conformal window. Thus studies of the $N_c=3,N_f=8,9$ theories, such as in 
\cite{LSD:2014nmn,LatticeStrongDynamics:2018hun} may encounter these problems, and indeed \cite{Hasenfratz:2022zsa} suggests that the artefact phase may lie close to the physical theory at $N_f=8$.
\vspace{-0.6cm}

\section{The Holographic Model}

Our simple holographic model consists of just a conformal {\it dimension 1} scalar in AdS$_5$. The setup shares many aspects of the top-down D3/probe D7 system \cite{Karch:2002sh,Kruczenski:2003be,Erdmenger:2007cm}. The action is
\begin{equation}
    S = \int d^4x d r ~ \frac{1}{2} r^3 (\partial_r \phi)^2 + r^3 V(\phi,r),
\end{equation}
where we only include a kinetic term in the holographic $r$ coordinate, since we will only consider the Lorentz invariant vacuum. The equation of motion is
\begin{equation}
    \partial_r ( r^3 \partial_r \phi) - r^3 \frac{dV}{d \phi} = 0.
\end{equation}
We set by hand
\begin{equation}
    \frac{dV}{d \phi} = \frac{1}{r^2} \Delta m^2[r^2 + \phi^2] ~\phi.
\end{equation}
If $\Delta m^2=0$, then the scalar describes a dimension 1 source and a dimension 3 operator (its solution is $\phi = m + c/r^2$). If $\Delta m^2$ is a constant, then it corrects the dimensions by $\gamma$ where
\begin{equation} \label{mg}
    \Delta m^2  = \gamma(\gamma-2).
\end{equation}
We allow $\Delta m^2$ to depend on $r$ to represent the holographic running of the theory. In addition we correct $r$ to $\sqrt{r^2 + \phi^2}$ which is dimensionally allowed but crucially plays the role that when $\phi$ grows from zero it can relieve a BF bound violaion in the deep IR, allowing for a stable solution for $\phi$.

As we have discussed above we will simply assume that
\begin{equation}
    \Delta m^2 = - k \alpha,
\end{equation}
where $k$ is a constant and $\alpha$ is the running coupling of the gauge theory which we extract from the lattice results for the SU(3) gauge theory with $N_f=10$ (Figure 1).

Note that in the perturbative regime at one-loop in the gauge theory $\gamma_{\rm pert}=3 C_2(R) \alpha/2 \pi $. For the fundamental representation $C_2(F)= (N_c^2-1)/2N_c$. For $N_c=3$ this gives $\gamma= 0.64 \alpha$.  In this regime, where $\gamma$ is small, we can expand (\ref{mg}) and find $\Delta m^2=-2 \gamma$, so $k= 3 C_2(R)/\pi$, which for $N_c=3$ gives $k=1.27$. This approximation would say that the $N_c=3,N_f=10$ theory would break chiral symmetry when $g^2=9.9$, before $g^2=15$.

On the other hand, if one wishes $\gamma=1$ to equate to the BF bound point when $\Delta m^2=-1$ then one would pick $k$ to be half this value, $k=0.64$. This would predict the $N_c=3,N_f=10$ theory would break chiral symmetry at $g^2=19$, above the $g^2=15$ fixed point identified on the lattice.

If one takes the lattice value of the fixed point for $\alpha =15/4 \pi$ and wanted $\Delta m^2=-1$ in the IR to enforce that it is the edge of the conformal window, then one would pick $k=0.84$.

The lattice data itself has $\gamma=0.6$ when $\alpha=15/4\pi$, and so suggests $\gamma=0.5 \alpha$, which is smaller than the one-loop result. The lattice analysis places the theory in the conformal window and would hence propose that $k < 0.84$. Our analysis is now to question whether the lattice might have misidentified the value of $k$ linking $\alpha$ and the BF bound violation, so that these values of $\gamma$ or $\alpha$ actually place the theory in the chirally broken phase.

Below, in theories that break chiral symmetry, we will call the scale at which the BF bound is violated $\Lambda_{BF}$. For very walking theories, this will essentially be the scale where the theory hits its IR fixed point.

N.b.: we can view changing $k$ for a fixed running profile as studying a range of theories near the conformal window edge - one is basically just assuming the coupling as a function of RG scale varies by an overall constant as one changes between theories. In the IR this is probably a reasonable ansatz for the $\alpha(\mu)$ function.

We first choose our running coupling function. We fit the lattice results \cite{Hasenfratz:2023wbr} for the beta function in the $N_c=3,N_f=10$ theory, constructing an interpolating function that we show in red in Figure 1. 
We then integrate to obtain the running $\alpha$ shown also in Figure 1. 

Now to find our vacuum solutions we solve the equation of motion with the on shell boundary conditions
\begin{equation}
    \phi (r_{\rm min})=r_{\rm min}, ~~~~~~~\partial_r \phi(r_{\rm min}) = 0.
\end{equation}
This boundary condition is motivated by that of similar D3/probe D7 system equations \cite{Alho:2013dka}. We interpret $L(r_{\rm min})$ as the IR constituent quark mass. We will consider cases where the choice of $k$ places this theory either

\begin{center}
\includegraphics[width=6.7cm,height=4.8cm]{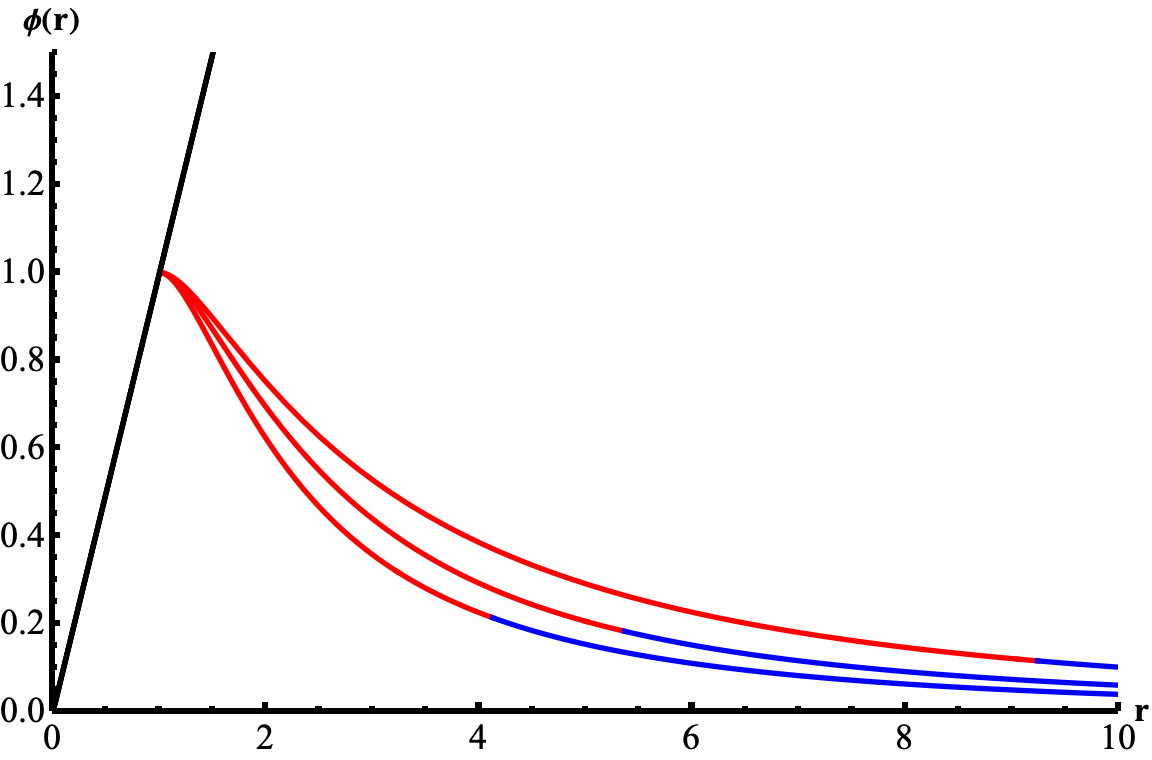}

\textit{Fig 2: The vacuum solutions $\phi$  for $k=3,1.8,1.4$ with the scale of $r_{\rm min}$ scaled to one in each case. The red part of each solution shows the $r$ range where the BF bound is broken at $\phi=0$. }
\end{center}

in the conformal window or in the chiral symmetry breaking phase above the conformal window edge.

We will now consider approaching the edge of the conformal window from above (the set of theories that break chiral symmetry) and from below (theories that are conformal).

\section{Approaching the edge of the Conformal Window from above}

First, let us consider the cases with $k>0.84$ - these theories will break chiral symmetry, approaching the edge of the conformal window as $k \rightarrow 0.84$.

In Figure 2 we plot the solutions of $\phi$ against $r$ for several $k$ values. For this plot, we have rescaled the energy scale of the theories so that the IR quark mass is equal in each case.  We have coloured the segments of the solutions (the range of $r$) where the BF bound is violated (at $\phi=0$) in red. This shows that at lower $k$ where the BF bound is violated less strongly, the theory takes a longer period in RG running to respond to the instability. In the limit where $k \rightarrow 0.84$ the gap between the BF bound violation scale and the IR mass scale diverges.

Note that if one increases $k$ too much, then chiral symmetry breaking is triggered in the far UV, where the running of the coupling is also slow. Here too, one sees a separation in scale between the BF bound violation point and the mass gap scale. Since these theories are breaking chiral symmetry in the very weakly coupled regime, we consider this unphysical. Thus, $k$ values above 3-4 are not sensible. 

To quantify the scale separation, we can define a scale where $\Delta m^2$  lies at $\Lambda_x = x \Lambda_{BF}$. In other words, if $x=0.95$ then it is the scale where the theory reaches a $\Delta M^2$ of 95\% of its value to cause a BF bound violation. In

\begin{center}
\includegraphics[width=6.7cm,height=4.8cm]{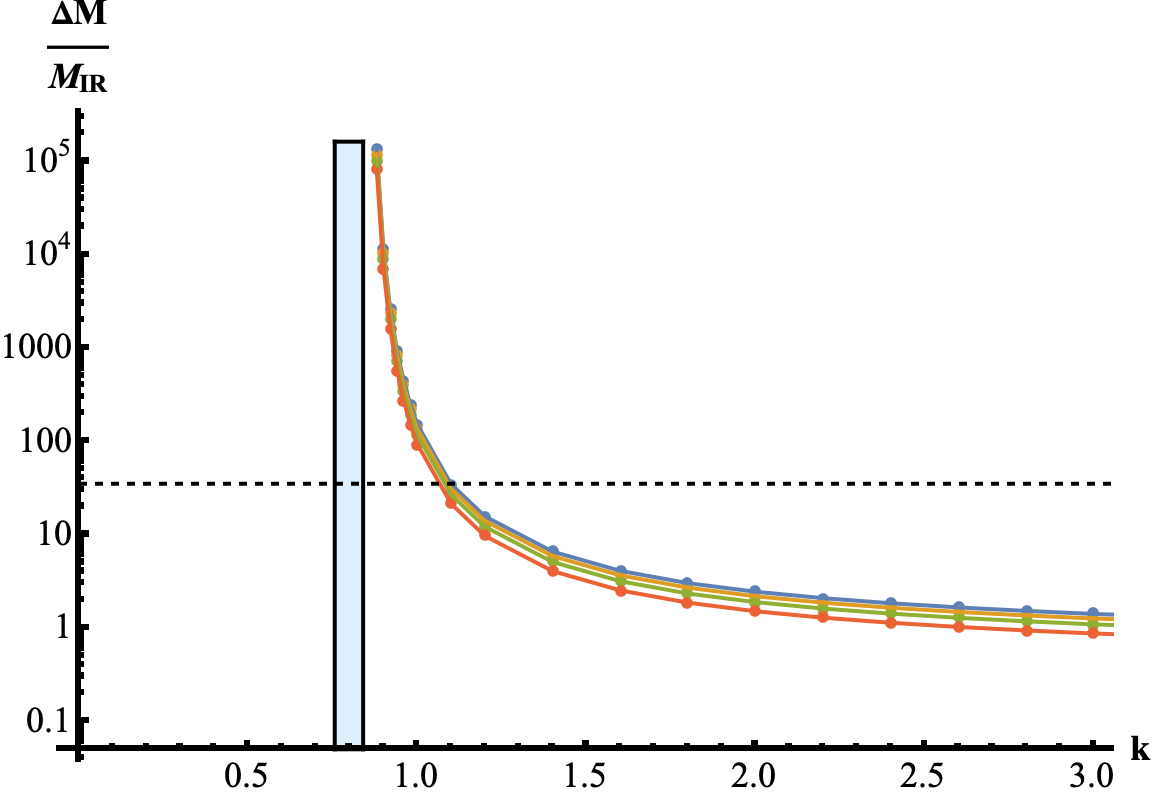}

\textit{Fig 3: Mass ratio as a function of $k$ for the four values of $x$ along with a 10\% area (blue box) under the minimum BF-bound violating value of $k=0.84$. Curves pass through a mass ratio of 35 at: $k=1.099$ for $x=0.8$, $k=1.088$ for $x=0.85$, $k=1.076$ for $x=0.9$ and $k=1.058$ for $x=0.95$. The blue shaded region below the critical $k$ value shows where the critical coupling lies within 10\% above the fixed point value. }
\end{center}

 figure 3 we plot the ratio of the IR quark mass scale to this scale for a variety of values of $x$. Note that the scale separation between this UV scale and the IR mass scale is greater than 35 if $0.84<k1.1$.

Let's put this in context for the lattice simulations: one might believe that one has a theory that lies in the conformal window and see fixed point behaviour. One could set the UV lattice coupling at 95\% of the fixed point value and use a $(35)^4$ point lattice and not see chiral symmetry breaking for these theories near the edge of the conformal window. However, we see that if $k$ lies in the range discussed, then it could be that just before the fixed point, an instability to chiral symmetry breaking sets in, yet the IR mass scale is below the IR resolution scale of the lattice.

Could this be the case for the SU(3) $N_f=10$ theory? If one reasonably thought the critical $k$ lies in the range 0.4-1.4 then 15\% of that range could actually be unseen chiral symmetry breaking if one didn't pick the UV coupling very carefully! In the next section, we will see that the work done with UV values of the coupling above the fixed point value strongly suggest this is not the case for that particular simulation. We will highlight additional concerns though, for theories nearer the conformal window edge.

\section{Approaching the edge of the Conformal Window from below}

If a lattice simulation of a theory suggests that a theory lies in the conformal window, then there is a simple proof of consistency. One studies the theory with UV lattice coupling value lying above the fixed point value. Such a theory in the continuum flows to the IR fixed point in the IR but will have a UV Landau pole. The UV cut off though, of the lattice provides protection from this pole and one can happily simulate. If such a case, with UV coupling above the fixed point, remains IR conformal then the evidence is overwhelming that the theory is indeed in the conformal window. This is the case for the SU(3) $N_f=10$ simulation \cite{Hasenfratz:2023wbr} and so its conclusions seem very robust.

Is this methodology always available though? It is well known that if on a lattice you raise the bare coupling too far, then a first order transition occurs to a new phase that is gapped on the scale of the lattice (and doesn't display rotational symmetry) \cite{Hasenfratz:2021zsl}. This phase is often called a ``lattice artefact phase". Here we want to connect that high coupling phase to the chiral symmetry breaking phase in the holographic model. We will argue that as one approaches the edge of the conformal window that phase lies very close to the true phase of the theory.

In the holographic model one can also naïvely study the theories with a Landau pole that flow to the IR fixed point from above. To exclude the pole one can include an explicit maximum value of $r$ and impose that $\phi$ should vanish there. Solutions of the form in Figure 2 exist - and continue to exist even as the UV cut off is moved close to the Landau pole. At some point as the UV cut of is raised, the UV value of $\Delta m^2$ violates the BF bound. In fact, even here there are solutions of the form in Figure 2 with the action choosing to minimize the $r$ derivatives in the IR as compensation for the UV BF bound violating potential term. These solutions should not be viewed as physical though. If there is a BF bound violation in the UV, there is an instability at that scale to the unlimited growth of $\bar{q} q$. The theory will gap at the UV scale and never reach the IR values of $r$.  If one tracks back the picture in Figure 2 to the D3/probe D7 model, then the statement is that there is a D7 embedding situated at the UV cut off and perpendicular to the $r$ axis which is energetically preferred (in fact, since the UV cut off there is for the radial distance $\sqrt{r^2 + \phi^2}$, the embedding is just a point and the theory is rather nonsensical). This transition, the instant the BF bound for $\phi$ is violated by the UV coupling, which is first order since the condensate jumps from zero to very large, seems analogous to the lattice artefact phase. Likely both the holographic and lattice model share this instability at the UV cut off scale. In the lattice case, since the regulator does not obey rotational symmetry, the instability may also be localized at some points on the lattice, but the spirit of the transition is shared.

There is a natural concern that arises from this discussion. At the edge of the conformal window, we expect the fixed point value of the coupling to just violate the BF bound. Were one to place the UV coupling to the fixed point value though, one would trigger a transition to the artefact phase. In other words, the artefact phase lies arbitrarily close to the physical phase as one approaches the conformal window.

To quantify this issue, we plot in Figure 3, as a band, the range of $k$ below $k=0.84$ in our theory where setting the UV coupling to 10\% above the fixed point value violates the BF bound. In this region a lattice simulation is going to have to tune the UV coupling above the fixed point to confirm the conformal nature of the fixed point. We can see from Figure 3 that theories with $0.76<k<0.84$ are very hard to study in this way - this is 12\% of the range $k=0.4-1.4$.

Let's assume the artefact phase is caused by a chiral instability as discussed: then we can use the fixed point values at $N_c=3,N_f=10$ ($g^2=15$ and $\gamma=0.6$) to estimate the coupling where $\gamma=1$, assuming they are proportional, to be $g^2=25$. This indeed broadly matches the value for the onset of the artefact phase seen in \cite{Hasenfratz:2021zsl}. 
An added complication on the lattice, that we will not study here, is that the rotational symmetry breaking nature of the lattice cut off will distort the nature of the transition to the artefact phase. The variation of the critical coupling to enter the artefact phase with different regulators is studied in \cite{Hasenfratz:2021zsl}. Hence there is an additional uncertainty in the precise value of the critical coupling in any lattice simulation.

\section{Discussion}

We have used a holographic model of chiral symmetry breaking in SU($N_c$) gauge theories with quarks in the fundamental representation to study the transition from the conformal window to the chiral symmetry breaking phase. We have used the running determined by the lattice simulations in \cite{Hasenfratz:2023wbr} at $N_f=10$ but included an additional parameter, $k$, that determines the relation between the anomalous dimension of $\bar{q}q$ $\gamma$ and the coupling $\alpha$.  Here, varying $k$ effectively moves us through the edge of the conformal window, playing the role of $N_f$.

There have been some recent lattice studies \cite{Hasenfratz:2022qan,Butt:2024kxi} that suggest that for particular $N_f$ values (e.g.: $N_f=8$ for $N_c=3$) an additional phase is possible where a four fermion operator condenses and the theory gaps without breaking chiral symmetry. This crucially depends on the absence of anomalies in these theories. Our model does not include this operator so we do not include this possibility. It would be interesting to provide a holographic model of this phase in future work.

For the transition from the conformal window to a chiral symmetry breaking phase, we have highlighted two issues. The first is that when one places the UV coupling below the fixed point value in chiral symmetry breaking theories, there are two scale separations near the edge of the conformal window. Firstly, there is a separation between the scale where the theory enters strong coupling and the scale where the instability to chiral symmetry breaking occurs. But in addition there can be a scale separation between the scale where the instability sets in and the actual scale of condensation. As one reaches the conformal window edge, both of these scale separations diverge.

The second issue is in theories that lie in the conformal window and where one simulates with the UV cut off coupling above the fixed point value. Here, the chosen UV coupling can lie above the critical coupling and this triggers an instability in the holographic model which we believe is equivalent to the lattice ``artefact phase" where chiral symmetry is broken at the lattice scale.  As one approaches the conformal window edge, this artefact phase lies arbitrarily close to the fixed point value.

We estimate that for theories within 10\% (above and below) of the conformal window edge, considerable tuning will be needed. Placing the UV coupling of the theory below the fixed point value is likely to not show chiral symmetry breaking even if it is present because of the scale separation. Moving the UV fixed point above the critical coupling will place the simulation in the artefact phase. One will be left tuning the UV coupling above the fixed point value seeking a lower edge to the artefact phase that may be very close to the fixed point value.

To finish on a more positive note though, our analysis does support the recent $N_f=10$ lattice analysis \cite{Hasenfratz:2023wbr}. There, a chiral symmetry preserving phase has been found for UV couplings above the fixed point value and that provides strong confirmation that the theory lies in the conformal window.

\bigskip \noindent {\bf Acknowledgements:}
The authors are grateful for discussions with Anna Hasenfratz. NE's work was supported by the STFC consolidated grant ST/X000583/1.

%%%%%%%%%%%%%%%%%%%%%%%%%%

\end{document}